\documentclass[
]{ceurart}

\usepackage{listings}
\usepackage{caption}
\usepackage{subcaption}

\usepackage{cleveref}
\begin{document}

\copyrightyear{2025}
\copyrightclause{Copyright for this paper by its authors.
  Use permitted under Creative Commons License Attribution 4.0
  International (CC BY 4.0). CEUR Workshop Proceedings (CEUR-WS.org)}

\conference{HCMIR25: 3rd Workshop on Human-Centric Music Information Research, September 20th, 2025, Daejeon, Korea}

\title{AImoclips: A Benchmark for Evaluating Emotion Conveyance in Text-to-Music Generation}


\author[1]{Gyehun Go}[%
orcid=0009-0006-2840-9643,
email=rotation@kaist.ac.kr
]
\address[1]{Korea Advanced Institute of Science and Technology (KAIST)}
\address[2]{Seoul National University (SNU)}

\author[2]{Satbyul Han}[%
orcid=0009-0000-8288-5671,
email=merrymu@snu.ac.kr
]

\author[2]{Ahyeon Choi}[%
orcid=0000-0001-8304-599X,
email=chah0623@snu.ac.kr
]

\author[1]{Eunjin Choi}[%
orcid=0009-0002-1930-6141,
email=jech@kaist.ac.kr
]

\author[1]{Juhan Nam}[%
orcid=0000-0003-2664-2119,
email=juhan.nam@kaist.ac.kr
]

\author[2]{Jeong Mi Park}[%
orcid=0000-0001-7172-686X,
email=mensa04@snu.ac.kr
]
\cormark[1]

\cortext[1]{Corresponding author.}


\begin{abstract}
Recent advances in text-to-music (TTM) generation have enabled controllable and expressive music creation using natural language prompts. However, the emotional fidelity of TTM systems remains largely underexplored compared to human preference or text alignment. In this study, we introduce AImoclips, a benchmark for evaluating how well TTM systems convey intended emotions to human listeners, covering both open-source and commercial models. We selected 12 emotion intents spanning four quadrants of the valence-arousal space, and used six state-of-the-art TTM systems to generate over 1,000 music clips. A total of 111 participants rated the perceived valence and arousal of each clip on a 9-point Likert scale. Our results show that commercial systems tend to produce music perceived as more pleasant than intended, while open-source systems tend to perform the opposite. Emotions are more accurately conveyed under high-arousal conditions across all models. Additionally, all systems exhibit a bias toward emotional neutrality, highlighting a key limitation in affective controllability. This benchmark offers valuable insights into model-specific emotion rendering characteristics and supports future development of emotionally aligned TTM systems.

\end{abstract}


\maketitle

\vspace{-0.5cm}
\section{Introduction}

Recent advances in generative AI have transformed creative domains, including music. AI music generation has evolved from unconditional \cite{oord2016wavenet, huang2018music, huang2020pop} to conditional generation \cite{roberts2018hierarchical, payne2019musenet, dhariwal2020jukebox, agostinelli2023musiclm} for better human controllability, with text-to-music (TTM) systems becoming prominent due to their familiar input format, user-friendliness, and expressive potential. Studies on the evaluation of TTM systems have also followed recent advances \cite{grotschla2025benchmarking, liu2025musiceval, huang2025aligning}, focusing on human preferences and text prompt alignment.

However, emotional aspects are still often overlooked in evaluations of TTM systems. Although several studies have addressed emotion-conditioned music generation, these have mainly focused on symbolic music \cite{hung2021emopia, ferreira2021learning, grekow2021monophonic, sulun2022symbolic, huang2024emotion}, and the question of how well TTM systems convey intended emotions to human listeners remains largely unexamined. Neglecting emotion evaluation can lead to mismatches between user expectations and generated music, limiting the use of TTM systems in real-world applications such as affective music recommendations or adaptive soundtracks. Thus, assessing the effectiveness of emotional communication is crucial for developing user-centered AI music systems.

In light of this background, we develop AImoclips, a benchmark that consists of AI-generated music clips for evaluating emotion conveyance of TTM systems. We generated 991 clips with six state-of-the-art models, and collected valence–arousal ratings from 111 human participants. We open-source the dataset, annotated with valence and arousal scores averaged across multiple raters. The complete dataset is open-source and available at \url{https://github.com/HunRotation/HunRotation.github.io}, with audio samples provided at \url{https://hunrotation.github.io/projects/aimoclips.html}. We also analyze the results to understand the characteristics of TTM systems regarding emotion conveyance, focusing on three main questions: (1) How differently do TTM systems convey intended emotions to humans? (2) What are the valence–arousal features of well- and poorly conveyed emotions? (3) How emotionally diverse are the clips generated by TTM systems?

\section{Related Work}
Music emotion analysis has led to the development of several publicly available datasets with emotion annotations. These include categorical \cite{hung2021emopia, turnbull2007towards, wang2014towards, YM2413-MDB} and dimensional approaches \cite{ferreira2021learning, soleymani20131000, aljanaki2017developing, speck2011comparative}, often using valence–arousal models. Recent benchmarks such as GlobalMood \cite{lee2025globalmood} address the cross-cultural differences in music emotion recognition.

In parallel with these developments, evaluating emotional fidelity in AI music generation has gained importance. While many studies rely on automatic metrics due to the cost of human evaluation, recent works have begun incorporating human preferences. However, most research in this area still focuses on assessing automatic metrics \cite{grotschla2025benchmarking} or improving them to better align with human judgments \cite{liu2025musiceval, huang2025aligning}.

Although relatively few studies directly address emotional conveyance, there have been efforts to analyze emotion conveyance in AI-generated music. The work most similar to ours is that of Gao et al. \cite{gao2024aiemotion}, who evaluated emotion conveyance accuracy in the valence–arousal plane using discrete emotion labels. In contrast, our study uses continuous valence–arousal ratings, involves significantly more clips and subjects, and includes a broader set of systems—both open-source and commercial—providing a more comprehensive evaluation of current TTM systems.

\section{Dataset Construction}

\subsection{Emotion Intents}
First, we selected 12 emotion words to represent emotional intents when generating music clips. We used two sets of words as candidates. The first set consists of 28 sample words that Russell \cite{russell1980circumplex} used to represent the domain of affect. The second set comprises adjectives that do not lie exactly on the valence or arousal axis, from the 12-Point Affect Circumplex Model (12-PAC) \cite{yik201112pac}. From the candidate words, we selected those included in the Warriner et al.'s extension \cite{warriner2013norms} of Affective Norms for English Words (ANEW) \cite{bradley1999affective} and filtered out words with valence or arousal scores between 4.0 and 6.0 on a 1.0–9.0 scale. This step was intended to exclude words with intermediate valence or arousal values, thereby clearly separating the groups of emotion intents corresponding to each quadrant of the valence-arousal plane. Finally, from the remaining words, we selected three for each quadrant, using 5.0 (both for valence and arousal) as the boundary. The selected words are listed in Table \ref{tab:emotionwords}.

\begin{table*}[htb!]
  \vspace{-0.2cm}
  \caption{Words chosen for emotion intent of TTM generation}
  \label{tab:emotionwords}
  \begin{tabular}{cccl}
    \toprule
    Quadrant&Valence&Arousal&Words\\
    \midrule
    Q1 & High & High & happy, excited, energetic\\
    Q2 & Low & High & angry, anxious, scared\\
    Q3 & Low & Low & sad, gloomy, dull\\
    Q4 & High & Low & relaxed, calm, tranquil\\
  \bottomrule
\end{tabular}
\vspace{-0.4cm}
\end{table*}

\subsection{TTM System Selection and Clip Generation}

Next, we selected six TTM systems, then generated 14 clips for each emotion intent. There are 12 emotion intents, resulting in 168 clips created by each system, forming a dataset of 1,008 AI-generated music clips. All clips were 10 seconds long. 

We utilized four open-source models—AudioLDM 2 \cite{liu2024audioldm}, MusicGen \cite{copet2023musicgen}, Mustango \cite{melechovsky2023mustango}, and Stable Audio Open \cite{evans2025stableaudio}—and two commercial models, Suno v4.5 \cite{Suno} and Udio v1.5 Allegro \cite{Udio}. AudioLDM 2 and Stable Audio Open are diffusion-based generative audio models \cite{ho2020denoising}, while MusicGen consists of an autoregressive transformer-based decoder \cite{vaswani2017attention}, and Mustango is a hybrid model based on both FLAN-T5 \cite{chung2024scaling}, an instruction-tuned language model, and a diffusion model. Detailed information on each model we used is summarized in Table \ref{tab:TTMmodels}, along with the sample rate we used. As the sample rate of each model’s generation was fixed per checkpoint, we could not unify the sample rates.

We use the same prompt format for all generations. For a given emotion intent \textit{{word}}, the text input to the TTM system was “\textit{{word}}, instrumental". The reason we added ‘instrumental’ to the prompt was to ensure that only music clips without vocals should be generated, as vocal lyrics can influence the emotional expression of the clip. For Suno and Udio, there is an ‘instrumental’ toggle option to prevent the models from generating vocal tracks. Thus, the text input contains only the emotion intent itself without additional words, and the ‘instrumental’ option is enabled.

For open-source models, each clip was generated with a length of 30 seconds, after which a 10-second segment was randomly cropped from between the 15\% and 85\% marks of the full clip to exclude the intro and outro, which typically contain less information. It was impossible to control the exact length of the output generated by Suno and Udio, so we applied the same cropping procedure, with the only difference being that the full-length clips had variable lengths. The lengths of all generated clips were at least 30 seconds, which was sufficient to extract a 10-second segment.

\begin{table*}[htb!]
  \vspace{-0.2cm}
  \caption{Information of TTM systems to generate music clips}
  \label{tab:TTMmodels}
  \begin{tabular}{cccc}
    \toprule
    Model&Checkpoint (Version)&Availability&Sample rate(kHz)\\
    \midrule
    AudioLDM 2& audioldm2-music & open-source & 16\\
    MusicGen & musicgen-medium & open-source & 32\\
    Mustango & mustango & open-source & 16\\
    Stable Audio Open & stable-audio-open-1.0 & open-source & 44.1\\
    Suno & Suno v4.5 & commercial & 44.1 \\
    Udio & Udio v1.5 Allegro & commercial & 44.1\\
    
  \bottomrule
\end{tabular}
\vspace{-0.4cm}
\end{table*}

\subsection{Human Evaluation}

We conducted an online survey to collect valence–arousal ratings for the generated music clips. For each of 111 participants, 56 clips were randomly selected to ensure an average of over five ratings per clip. Each clip was selected from a distinct combination of TTM system and emotion intent, and presented in random order. Participants rated the conveyed valence and arousal on a 9-point Likert scale. All evaluation activites involving human subjects in this study were approved by the IRB at Seoul National University under approval number IRB No.2505/001-012.

Before rating the clips, participants were asked to provide demographic information, such as gender and age. This procedure ensured that all participants were fluent Korean speakers aged at least 18. They were then shown a sample question and clip to confirm that they could hear the audio and understand the instructions. The sample clip was neither rated nor included in the results. The only other data collected was contact information for the purpose of providing a participation reward.

\begin{figure}[htb!]
  \centering
  \includegraphics[width=0.5\linewidth]{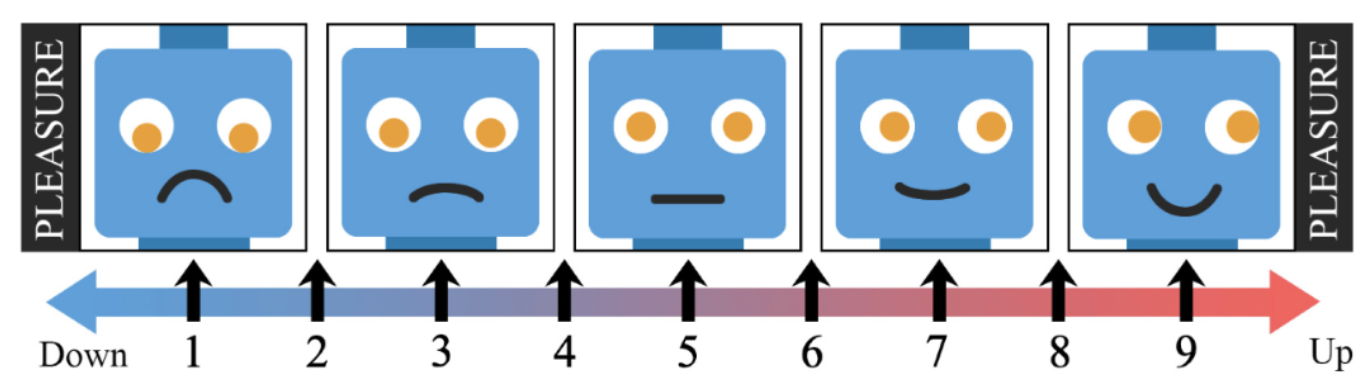}
  \includegraphics[width=0.5\linewidth]{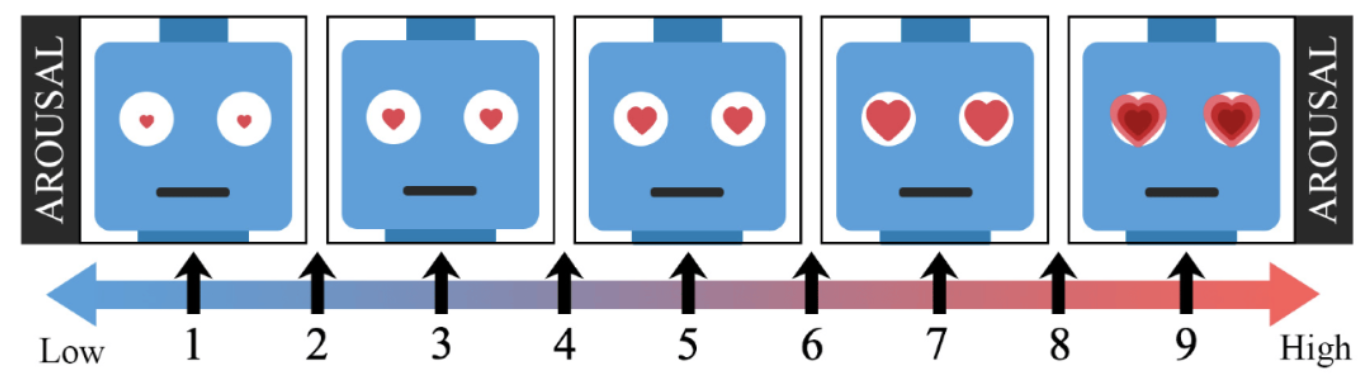}
  \caption{Survey images for valence (top) and arousal (bottom) rating questions. Adapted from He et al. \cite{he2023emotional}.}
  \label{fig:surveyimage}
\end{figure}

During the human evaluation conducted via the survey website, participants were presented with each clip one at a time and asked (in Korean): “Rate the pleasure conveyed by this music clip, with 1 meaning ‘very unpleasant’ and 9 meaning ‘very pleasant,’” and “Rate the arousal conveyed by this music clip, with 1 meaning ‘very relaxed’ and 9 meaning ‘very aroused.’” To ensure the concepts of valence (pleasure) and arousal were clearly conveyed, the images in Figure \ref{fig:surveyimage} were displayed alongside both questions. This method, adapted from He et al. \cite{he2023emotional}, combines visual aids with the explicit textual anchors provided in the questions. Participants selected integer scores from 1 to 9 for each question and were required to provide one rating per question to proceed to the next clip, similar to the static annotation method used in the DEAM dataset \cite{aljanaki2017developing}.

When selecting music clips for a new participant, priority was given to clips rated by fewer participants to help balance the number of ratings for each clip. However, it was difficult to completely balance the ratings, as some participants did not complete the survey. As a result, the number of participants who rated each clip ranged from 2 to 9, and we excluded 17 clips with three or fewer ratings, leaving 991 clips in the final dataset. Thus, the number of ratings on each clip ranges from 4 to 9.

\section{Results}

\subsection{Participants and Number of Ratings}

\begin{figure}[htb!]
  \centering
  \includegraphics[width=.40\linewidth]{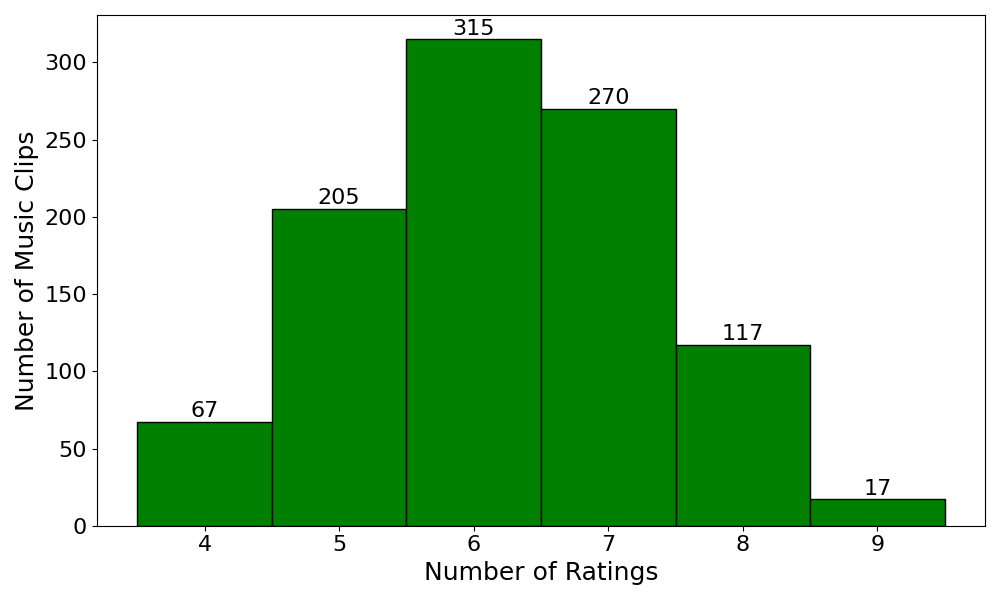}
  \caption{Distribution of music clips by number of ratings per clip.}
  \label{fig:appendix_ratingcount}
  \vspace{-0.4cm}
\end{figure}

111 participants rated the generated music clips, yielding 6,162 ratings after clips with three or fewer raters were excluded. For one participant, six ratings were removed because the participant was unable to listen to the last six samples due to a technical issue with the survey website. The participant group was composed of 47 males and 64 females. The majority (89 of 111) were between 18 and 29 years old, while the remainder included middle-aged participants, six aged 45–49 and five aged 50 or older. The detailed distribution of ratings per clip is presented in \ref{fig:appendix_ratingcount}.

\subsection{Overall Score Distribution for Each TTM System}

\begin{figure}[htb!]
  \centering
  \begin{subfigure}{.44\textwidth}
    \centering
    \includegraphics[width=.99\linewidth]{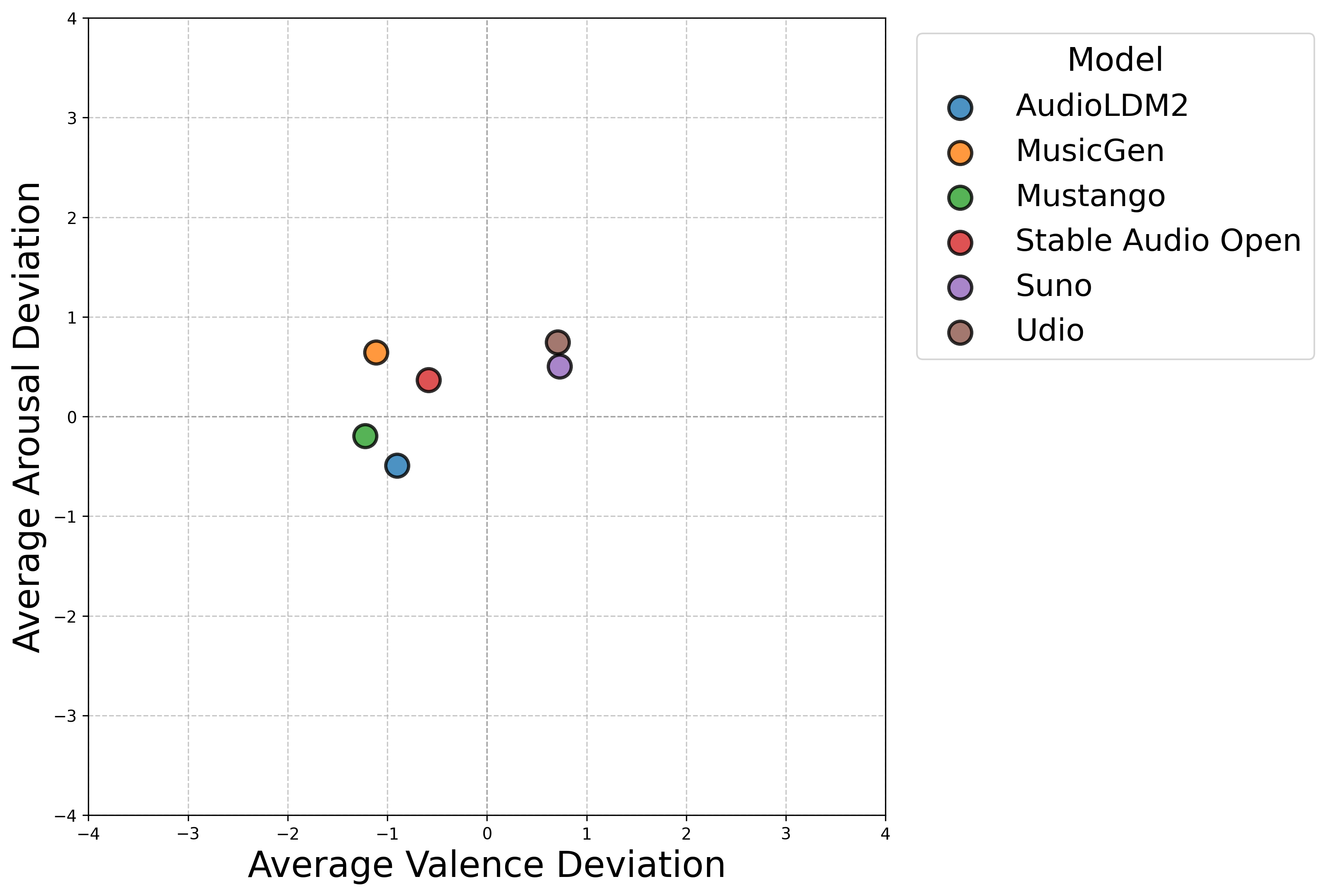}
    \subcaption{}
    \label{fig:model_avg}
  \end{subfigure}
  \begin{subfigure}{.46\textwidth}
    \centering
    \includegraphics[width=.95\linewidth]{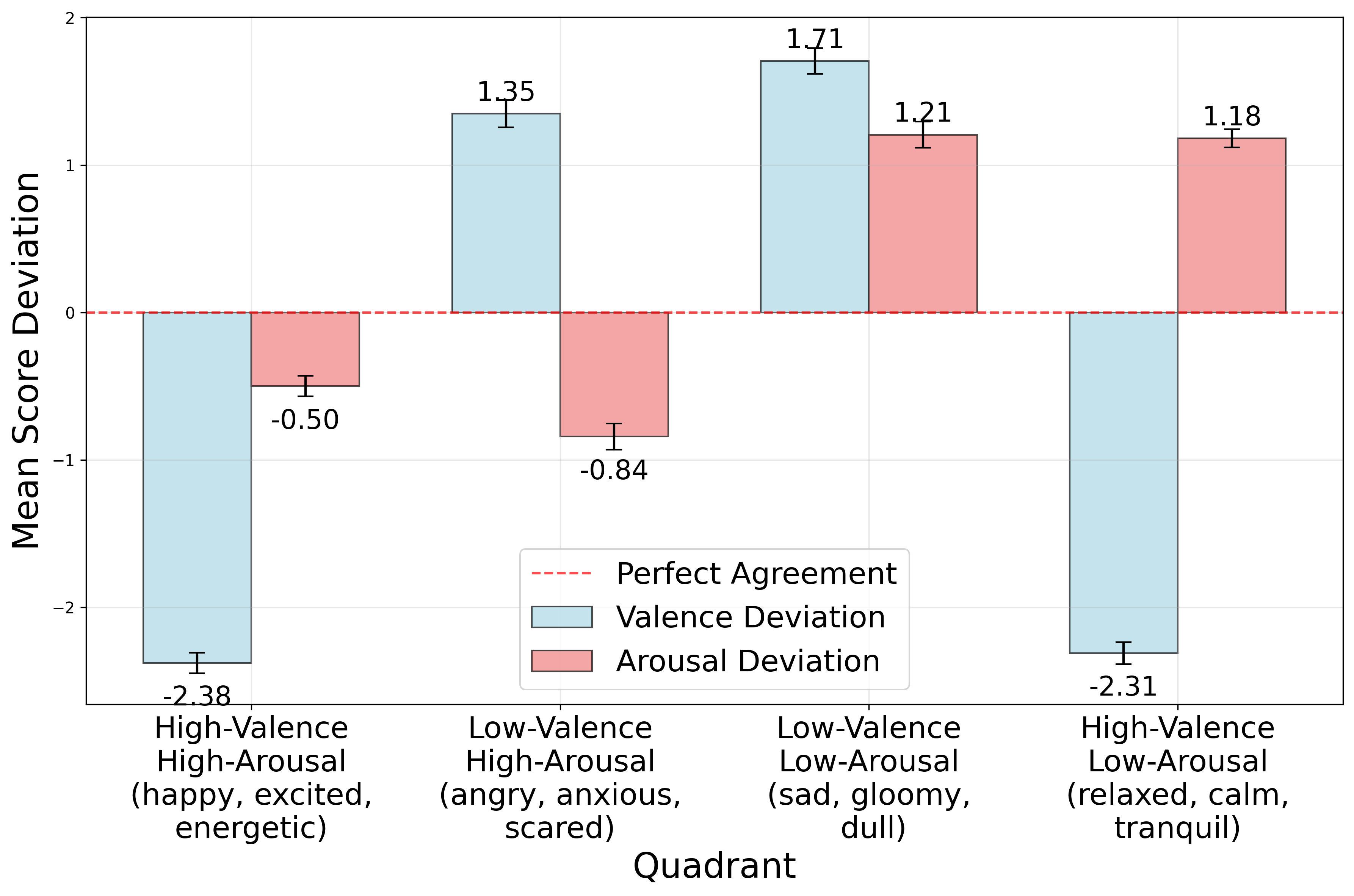}
    \subcaption{}
    \label{fig:quadrant_mean_comparison}
  \end{subfigure}
  \caption{(a) Mean valence and arousal deviations per model, computed by averaging (clip ratings - corresponding emotion intent scores). (b) Mean deviations per valence–arousal quadrant, aggregated across all models. }
  \label{fig:total_results}
\end{figure}

\begin{figure}[htb!]
  \centering
  \begin{subfigure}{.43\textwidth}
    \centering
    \includegraphics[width=.99\linewidth]{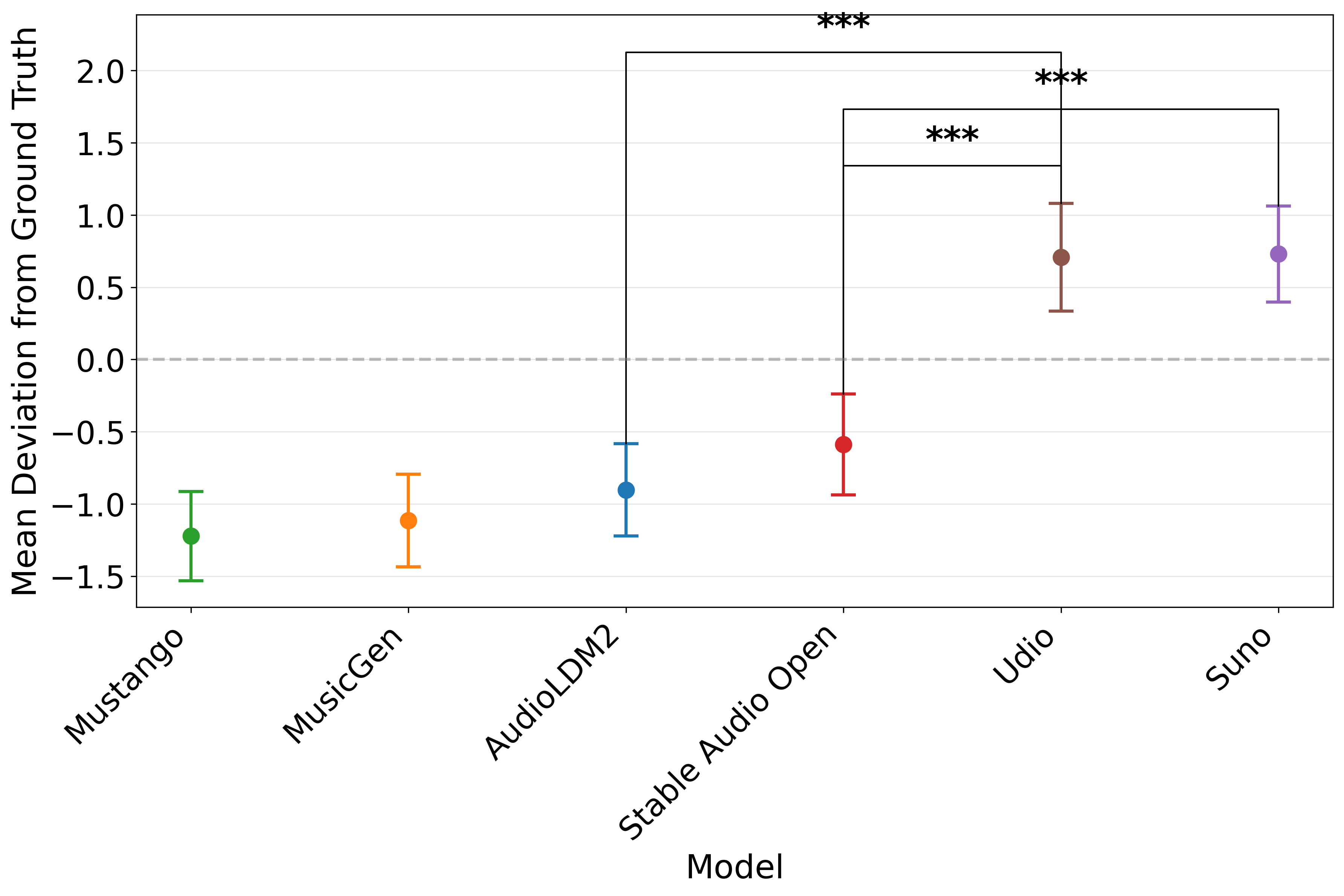}
    \subcaption{}
    \label{fig:appendix_model_valence_mean_comparison}
  \end{subfigure}
  \begin{subfigure}{.43\textwidth}
    \centering
    \includegraphics[width=.99\linewidth]{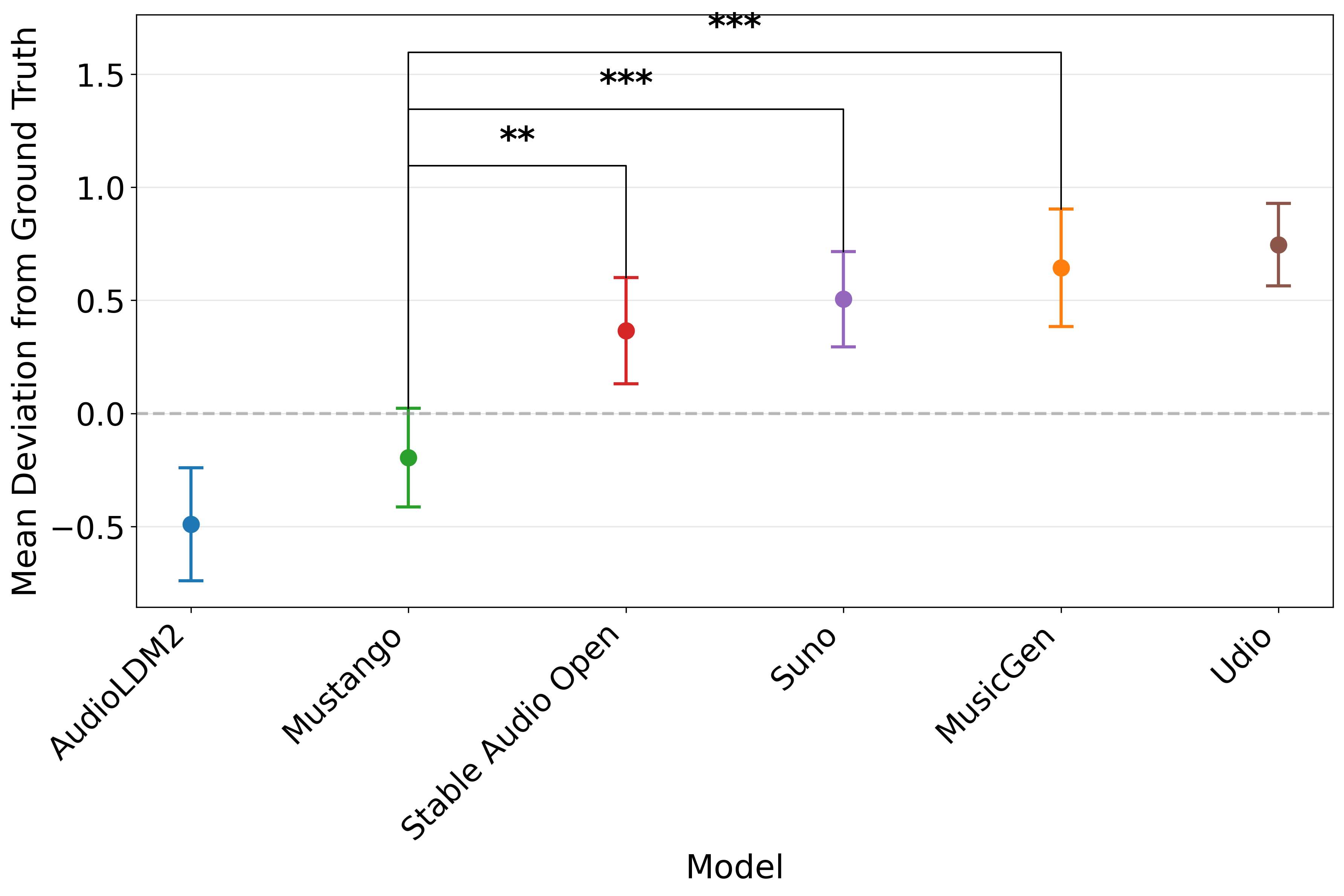}
    \subcaption{}
    \label{fig:appendix_model_arousal_mean_comparison}
  \end{subfigure}
  \caption{Mean deviation plots for each model with 95\% confidence intervals. Among all significantly different pairs, the three with the smallest differences are shown. (*: p<0.05, **: p<0.01, ***: p<0.001) (a) Mean valence deviations. (b) Mean arousal deviations.}
  \label{fig:appendix_model_mean_comparison}
\end{figure}

As shown in Figure \ref{fig:model_avg}, valence and arousal deviations from the intended emotion differ across models. We used scores from Warriner et al's dictionary \cite{warriner2013norms} as the ground truth. Open-source models tend to produce music perceived as less pleasant than intended, resulting in lower valence ratings compared to the ground truth. In contrast, generations from Suno and Udio yield higher valence ratings, forming a clear separation between open-source and commercial models. Arousal deviations are less distinct but still form two clusters: AudioLDM 2 and Mustango show negative arousal deviations, while the others show positive. A two-way ANOVA using TTM system and emotion intent as factors confirms significant model effect for both valence($F_{(5, 974)} = 150.53, p<.001, \eta^2 = 0.123$) and arousal($F_{(5, 974)} = 39.44, p<.001, \eta^2 = 0.0836$) deviance. Moreover, pairwise comparisons (Figure \ref{fig:appendix_model_mean_comparison}) support these clusters, showing significant differences between but not within clusters.

\begin{figure}[htb!]
  \centering
  \includegraphics[width=.99\linewidth]{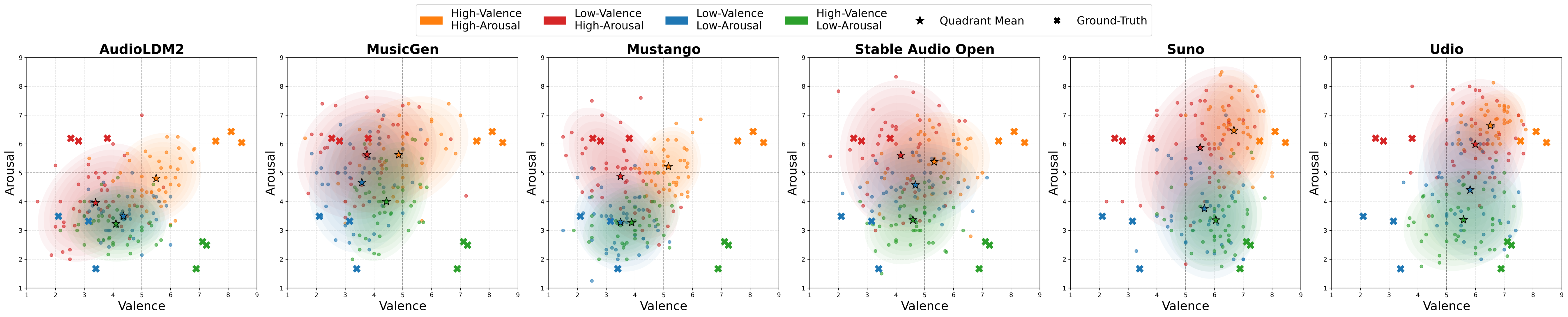}
  \caption{Valence–arousal quadrant distributions by model. Stars show mean ratings per quadrant, 'X' marks represent ground truth scores of emotion intents, and ellipses indicate 95\% confidence regions.}
  \label{fig:quadrant_scatter}  
  \vspace{-0.3cm}
\end{figure}

Figure \ref{fig:quadrant_scatter} shows the distribution of valence-arousal ratings for clips generated by each model. For all models, the clips tend to be rated more neutrally than their original emotion words in text. Ratings for clips generated by Suno and Udio are biased toward higher valence, indicating that these models tend to generate music conveying more pleasant emotions. In contrast, the rating distributions for open-source models tend to lean toward lower valence, with AudioLDM2 and Mustango also exhibiting a bias toward lower arousal. Another notable observation is that Suno and Udio display vertically wider distributions, suggesting that they generate music clips covering a broader range of arousal.

\subsection{Quadrants in the Valence-Arousal Plane}

\begin{figure}[htb!]
  \centering
  \includegraphics[width=.66\linewidth]{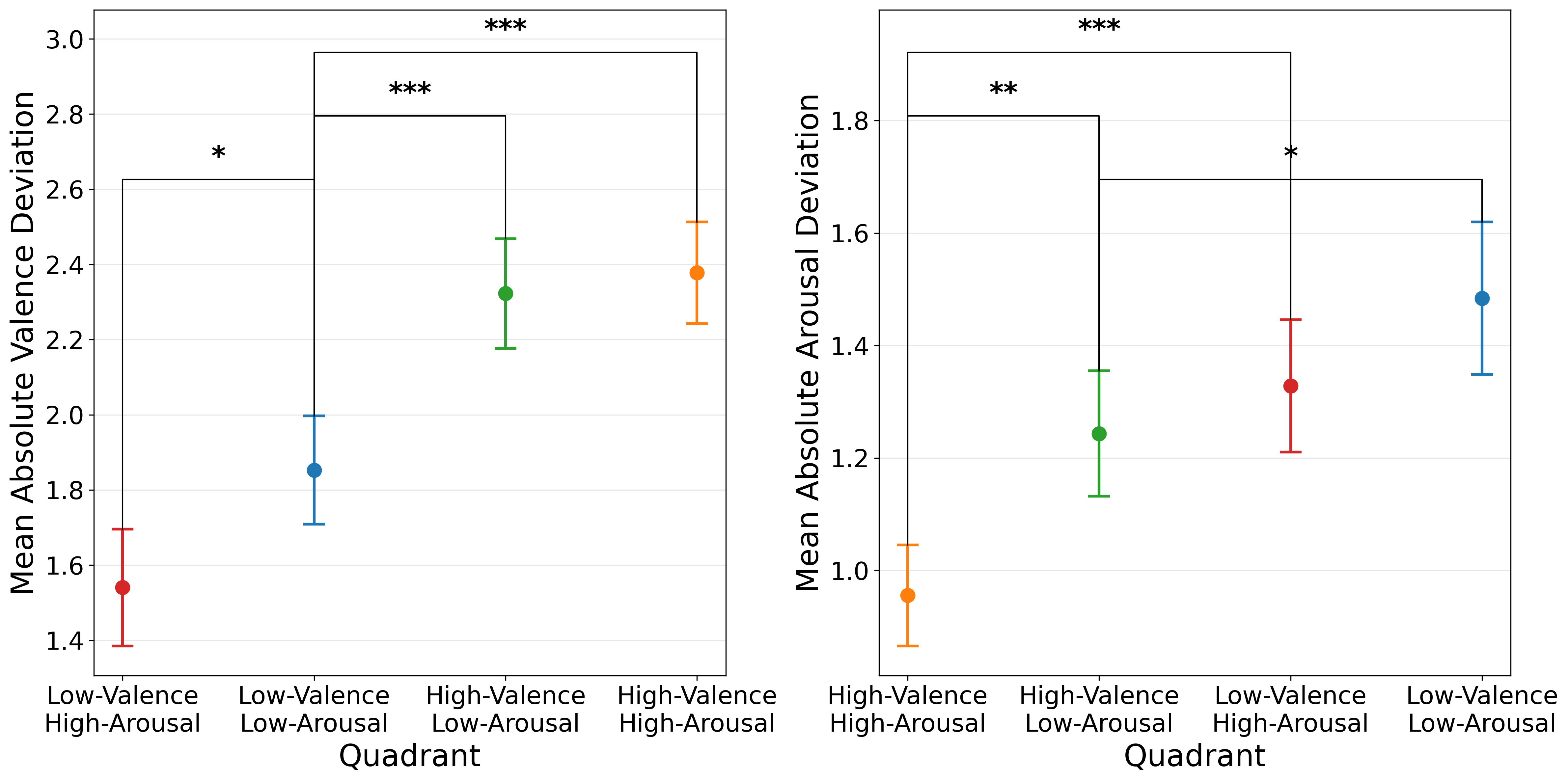}
  \caption{Mean absolute deviation plots for each valence-arousal quadrant with 95\% confidence intervals. Among all significantly different pairs, the three with the smallest differences are shown. (*: p<0.05, **: p<0.01, ***: p<0.001) (a) Mean absolute valence deviations. (b) Mean absolute arousal deviations.}
  \label{fig:appendix_quadrant_absolute_mean_comparison}
  \vspace{-0.3cm}
  
\end{figure}

To analyze which emotions are well or poorly conveyed in the valence-arousal plane, we divided emotion intents into four quadrants and compared how closely the ratings of clips in each quadrant matched the ground truth scores. Figure \ref{fig:quadrant_mean_comparison} shows that valence is most accurately reflected when the intended emotion has low valence and high arousal—such as \textit{angry, anxious } and \textit{scared}. For arousal, emotion intents with both high valence and high arousal—such as \textit{happy, excited } and \textit{energetic}—are best conveyed. A two-way ANOVA with the quadrant and TTM system as factors confirms a significant main effect of the quadrant on absolute deviations for both valence($F_{(3, 982)} = 29.27, p<.001, \eta^2 = 0.0814$) and arousal($F_{(3, 982)} = 14.62, p<.001, \eta^2 = 0.0421$). In addition, the quadrants with the smallest magnitude of deviations show statistically significant differences from the others (Figure \ref{fig:appendix_quadrant_absolute_mean_comparison}). The differences between the emotion word scores and the corresponding clip ratings are larger for valence than for arousal, suggesting that the models capture intended arousal more successfully than intended valence.

Looking at the confidence regions displayed in Figure \ref{fig:quadrant_scatter}, it is notable that the distributions of quadrants with different arousal levels are well separated, whereas those with different valence levels show substantial overlap. In particular, when focusing on the two regions with low arousal level (green and blue regions), valence is not clearly distinguished when arousal level is low. Additionally, for Suno and Udio, the region corresponding to clips generated from high-valence, high-arousal intents is largely overlapped by the region for clips generated from low-valence, high-arousal intents.

\section{Discussion}

AImoclips is a large-scale benchmark with continuous emotion annotations for evaluating emotional fidelity in TTM generation, incorporating both open-source and commercial systems. By collecting valence–arousal ratings from human listeners and analyzing them statistically, we provide insights into model-specific biases and the current limitations in affective controllability.

A key finding is that all systems exhibit a centralizing tendency toward emotional neutrality, implying a limited ability to express subtle or polarized affective states. The result shows that, to human listeners, music from TTM systems expresses valence and arousal traits less clearly than textual expressions.

Our findings also highlight how differently current TTM systems convey emotional intent. Commercial models (Suno and Udio) tend to generate music perceived as more emotionally engaging than intended, while open-source models often produce less pleasant outputs. One plausible explanation is that listeners perceive more pleasant emotions in music they prefer or in high-quality audio. As Suno and Udio generally receive higher preference ratings \cite{grotschla2025benchmarking} and feature higher sample rates than the open-source models (Table \ref{tab:TTMmodels}), their outputs may be perceived more pleasant.

In terms of emotion types, high-arousal emotions such as \textit{excited} and \textit{angry} are more reliably expressed than low-arousal emotions like \textit{calm} or \textit{gloomy}. This partially aligns with Gao et al. \cite{gao2024aiemotion}, who noted that AI-generated music is particularly effective at conveying high-valence, high-arousal emotions.

Future research should investigate the specific acoustic and musical features that cause biased valence and arousal perception in AI-generated music. A primary focus should be on identifying the key characteristics of both TTM systems and human auditory perception to improve the alignment between generative intent and listener experience. The AImoclips benchmark can directly support this work by serving as a dataset for a variety of applications, such as training predictive models of human emotion ratings or fine-tuning generative models for enhanced affective controllability.


\section*{Declaration on Generative AI}

The authors used GPT-4.5 and Gemini 2.5 Pro during the preparation of this work for grammar and spelling checks only. All content was subsequently reviewed and edited by the authors, who take full responsibility for the final manuscript.

\bibliography{main}

\end{document}